\documentclass[12pt]{article}
\usepackage{psfig}

\begin{document}
\begin{center}

{\Large \bf Stellar disks and halos of the edge-on spiral galaxies: NGC~891,
NGC~4144 and NGC~4244}

\bigskip

{\large \bf Tikhonov N.A.
and Galazutdinova O.A.}

\bigskip

{\large \em Special astrophysical observatory RAS}

\bigskip
{\large \bf Abstract}
\end{center}

The results of the stellar photometry of the images ACS/WFC and WFPC2 of
the HST
 are used to study stellar population and spatial distribution of stars
in three
edge-on galaxies: NGC~891, NGC~4144 and NGC~4244.
The measuring of the number density of the old stars revealed two stellar
substructures in these galaxies: thick disk and halo. The borders of these
substructures consisting mainly of red giants, are determined by the change
of number density gradient of the old stars. The revealed halos have flattened
shapes and extend up to 25 kpc from the galaxy planes. The obtained results
 of
number density distributions of different type stars perpendicular to the
galaxy
planes allow us to verify our stellar model of spiral galaxies. Using the
determination of the tip of red giant branch (TRGB method) we have derived
 the
following distances: $D$ = 9.82 Mpc (NGC~891), $D$ = 7.24 Mpc (NGC~4144)
$D$ = 4.29 Mpc (NGC~4244).

\section{Introduction}
  By the present time the stellar population of galaxies has been
studied in sufficient details only  in the nearest  stellar systems
of the Local group where three spiral galaxies:
M~31, M~33 and Galaxy dominate. In the named of spiral galaxies
four stellar subsystems are generally distiguished: a bulge, a thin disk,
a thick disk and a halo. These subsystems have different laws of spatial
distributions of stars and consist of  stars and clusters
with a different average age and the mean metallicity value [1--6].
Due to the  fact that these subsystems are  embeded one into
another, it is difficult to define the space  boundaries of these
subsystems since the inner subsystem also contain stars of the outer
subsystems. The division of stars into subsystems is usually performed
on the basis of measurements of photometric parameters of stars, their
metallicity value and kinematic characteristics. The technical difficulties
at the vast determinations of characteristics of stars and the uncertainty
in the terminology when describing stellar morphology of galaxies cause some chaos
in the designation of the studied stellar structures of galaxies.
The circumstance that observations of galaxies are carried out in
different spectral bands, which naturally leads to changes in the
visible morphology of galaxies, makes the chaos  worse. On the basis of
surface photometry methods the procedure of division of spiral galaxies
into two components: a bulge and disk [7] has been developed reliably
enough. By the term ``disk'', a thin disk, where star formation is
currently in progress,
and the bright part of a thick disk inhabits by older stars
is implied. As far as a stellar halo of galaxies is concerned neither
the shape of the halo nor the law of the decrease of number density
toward the outer part of the halo are known yet.

By analogy with spiral galaxies of the Local group search for thick disks
and
halos are being carried out in more distant galaxies. For instance, on
the basis of studying of 47 late types galaxies it has been found that
most of these galaxies possesses a thick dusk [8]. However, this is more
likely
to be a qualitative pattern of the structure of galaxies. Since for revealing
thick disks methods of surface photometry were used, the determination of
 age
composition of stellar disks and establishment of their borders are
extremely involved problems. The situation with the search for and
determination of halos turns out to be even more intricate. In a number of
papers the term ``halo'' is used to name  any stellar structure beyond the
border  of thin disk. In this case the term ``thick disk''
is absent and its place is taken  by the term ``halo''. Naturally,
such a pseudo-halo can be recorded in some galaxies by the methods of
deep surface photometry.
For the theories of origin and evolution of galaxies it is extremely
necessary to have uniform and reliable results on extended stellar
structures of galaxies since they mostly consist of old stars with
low metallicity and keep information about the first stages of evolution
of
galaxies. Many results on stellar composition of nearby galaxies have
already been obtained, but the main step should be made and define the
conception
disks and halos. Without such definitions continuous confusion will take
place
in designations of those stellar structures which are observed in galaxies
with using various techniques.
Irregular galaxies, which are the close relatives of spiral galaxies,
have the simplest composition. The discovery of a stellar halo in two
irregular galaxies, WLM and NGC~3109, was announced in the papers
by Minniti and his colleagues [9], [10]. Later,  based on the study
of the stellar population of 25 irregular galaxies, Tikhonov [11-13]
showed that every irregular galaxy has a thick disk the dimensions of
which are 2--3 times  as large as the visible body of the galaxy. Young
stars in irregular galaxies concentrate toward the galactic plane and
form a disk similar to thin disks of spiral galaxies.
The study of massive irregular galaxies has shown that besides a thick
disk they may also have a halo that extends to a distance greater
than the thick disk and has a different gradient of decrease in stellar
number density. By the present time, we have found such halos around
a few massive irregular galaxies: IC~10, M~82, NGC~3077. On the basis of
these results it becomes clear that Minniti and his colleagues are likely
to have found in the galaxies WLM and NGC~3109 thick disks, but not
halos, since they have not found bend of the density gradient of red giants
and the sharp drop of stellar density at the  border of the thick disk
as
it is seen in the galaxy IC~10 [14], [11] and M~82.
Three spiral galaxies, NGC~55, NGC~300 and M~81, that we investigated earlier
show the presence of a thick disks and a halos
[15] which unites its morphologically with massive irregular galaxies.
In order to understand the  spatial structures of thick disks and halos, it
is necessary to investigate edge-on galaxies since only with such
position of galaxies we can see and
study the extent of the thick disk and halo perpendicular to the galactic
plane. In the present paper we fill the existing gap and present the
results of investigation of the stellar population of three edge-on galaxies.
The basic characteristics of these galaxies
are listed in Table 1. All the galaxies have a developed spiral structure,
and their luminosity is comparable with the luminosity of our Galaxy.

\subsection{NGC~891}. The galaxy NGC~891 belonging to the NGC~1023 group
is morphologically the most identical to our Galaxy. It might be expected
that it also has  similar physical conditions, however, the current rate of
star formation in NGC~891 is 2--3 times as high as that in our Galaxy.

The age of NGC~891 is estimated to be 11--13 billion years [16], [17],
which within the errors, corresponds to the age of our Galaxy equal to
13.6 billion years [18]. NGC~891 has been intensitively studied in all
spectral bands. The  investigation of optical images (BVI+H$\alpha$)
of NGC~891 shows two physically different components of the interstellar
disk of this galaxy: dense cold one  visible on BVI images in the form
of absorption and consisting of structurized dust clouds
traced to 2 kpc from the galactic plane and a warm
ionized one, visible on H$\alpha$ images and uniformly distributed
over the body of the galaxy with inclusion of thread-like structures
[19]. Photographic observations of NGC~891 on red and blue plates has revealed
two stellar components --- a bulge and a disk [20], the border of the disk
was traced almost to the border of the thick disk that we have
found on the basis of searching of stellar number density.
Radio investigations of NGC891 have shown that the galaxy has a gaseous
halo, where one observed CO [21] and atomic hydrogen [22], extending to
5 kpc from the galactic plane and being of $\sim$15\% of the total mass of
neutral hydrogen. The most remote HI clouds have been recorded to a
distance of up to 15 kpc from the galactic plane [23],[24]. The
diameter of the hydrogen disk of NGC~891 is 1.2 times as large as the
optical diameter of the galaxy determined from the level of the isophote
$\mu=25^m$/sq.arcsec in the  $B$ band
[25].
When using a axially-symmetrical model of the distribution of stars and
dust, three parameters were determined for NGC~891, which describe best
the distribution of stars and dust in the optical and IR regions [26].
The scale height in this model is equal 0.4 kpc for the stars and
0.26 kpc for the dust. The model gives the mass of dust
1.14$\times10^8M_{\odot} and M_{gas}/M_{dust}$=1.65, which is close to
the value obtained for our Galaxy [27].

\subsection{NGC~4144} Due to its high distance and absence of any morphology
peculiarities, the galaxy NGC~4144 becomes most frequency as an object
of statistical investigations of galaxies only [28-30]. A comparison
of HI radio observation and  optical observations in the $B$ and
 $R$ bands has  shown that the galaxy possesses a thick hydrogen disk
whose dimensions exceed those of the visual body of the galaxy [31], [32].

\section{NGC~4244} The galaxy has a low of star formation rate and is
surprisingly quest in the radio continuum [33]. Observations of atomic
hydrogen show that on the Z axis hydrogen is seen up to a height of $2.5
^{\prime}$
(3.1 kpc) [34]. The diameter of the hydrogen disk is 1.3 times larger than
the optical diameter at $\mu=25^m$/sq.arcsec
in the $B$ band [25]. The galaxy is of
low luminosity in the filter H$\alpha$ and of low surface brightness in
the far IR [35]. The deep CCD photometry (up to 27.5 stellar
magnitudes in the $R$ band) shows that NGC~4244 has a simple structure:
an exponential disk with a scale height of 250 pc. No evidence of
existence of the second component, a thick disk or a halo, has been
found [36].

\section{Observations}
 To study the stellar population of the galaxies, we used the archive images
obtained in   different years with the ACS/WFC and WFPC2 of the
Hubble Space Telescope (HST). The information about the archive images
that we have used is in Table 2.

Figs.1, 2 and 3 present the DSS images of the investigated galaxies
with ACS/WFC and WFPC2 footprint overlaid.
The stellar photometry was performed with the packages
of programs DAOPHOT II in MIDAS and HSTPHOT [37], [38]. The
reduction of DAOPHOT instrumental stellar magnitudes
to the standard Kron-Cousins system
was executed by the procedure described in the papers by
Holtzmann et al. [39],[40]. The reduction of the ACS/WFC results to
the standard VI system was performed on the basis of calibration
relationships for the stars in the galaxy IC~10 where we
fulfilled photometry of the same stars as both in ACS/WFC and in the
WFPC2 frames. To obtain colour relationship 60 calibration stars with
0$ < (V-I) <5$ were employed. The accuracy of the reduction of stellar
magnitudes is equal to 0.03$^m$ for the $I$ band and 0.04$^m$ for the
$V$ ones.

The derived equations of the reduction of instrumental stellar magnitudes
into the VI Kron-Cousins system were checked by us at the stars of
the known galaxies (NGC~55, NGC~300) and showed a good agreement of the
obtained results with the results of other authors. Background distant
galaxies, unresolved stars due to superposition of images of
close neighbors and stars ruined because of the defects of the CCD chips
were excluded from the final list on the basis of comparison of their
profiles with the standard star PSF profile.  The stars that
were left in the final lists had the following parameters:
``SHARP'' $<$ 0.3, and ``CHI'' $<$ 1.2 [37].

\section{Selection of stars}
To ensure maximum homogeneity of the sample of stars of different fields,
we made selection of stars basing on the following conditions:
a) To exclude the effect the duration of the exposure has on the number
of stars found in the field (images of different research programs were
used for different fields) we used the results obtained with the shortest
exposure. For all the studied
fields of one galaxy we set the same border in luminosity of
stars. Stars more fainter were not used even if they were seen
in the images of some fields of the galaxy.
b) The process of filtering of traces of cosmic particles removed them
to a considerable degree, and the particle could not distort the
counts of stars number density in crowded stellar
fields. However, in sparse stellar fields the number of unremoved traces
of cosmic particles (i.e. fictitious stars) is comparable with the number
of real faint stars. This effect leads to an imaginary decrease
in the gradients of star number density in the searching fields. In order to
eliminate this effect, we had to use stars by $0.5^m$ stellar magnitude
brighter than the photometric limit of the images since the program of
filtering particles is unable to remove the faintest traces of
particles. Such constraints reducted the number of stars in the fields
and increased  statistical noise, but to make up for it, we could be
certain that our results of measuring the number density of stars are not
affected by the residual traces of cosmic particles.

\section{Results of photometry}

\subsection{``Colour--magnitude'' diagrams of the galaxies: NGC 891,
NGC 4144, NGC 4244.}

The results of stellar photometry are presented in
Fig.4 in the form of ``Colour -- Magnitude'' diagrams (CMD). The obtained
diagrams are typical of spiral galaxies. The considerable differences
in the diagrams of different fields of one and the same galaxy can readily
be explained since the fields under investigation may cover both the
regions of spiral arms with bright supergiants and the regions of halos
where AGB stars can be the brightest stars, but the main population visible
in the images is the red giants.  Lines are drowns on all the
diagrams, which show the position of the tip of the red giant branch
 (TRGB) used for measurement of distances. In the field S2 of the galaxy
NGC~891, which is the most poorly populated by stars, only the tip of the
red giant branch is visible at (V-I) = 1.5 and a comparable number of
background stars.
Due to the closeness of the photometric limit of the images of this fild to
the luminosity of the searching stars, we moderated our criteria of selection
of red giants for this unique case, following from the real assumption
of large photometric errors of these faint stars. Our criteria for this
field are:  $1.0   < (V-I)<2.2$,$ I> 25.7$.

\subsection{Measurement of distances}
The presence of stars of the red giant branch on the ``Colour--Magnitude''
diagrams that we derived (Fig.4) enables to measure accurate distances
to the galaxies by using the TRGB method [41].

To calculate the distances, we used one or two fields in each galaxy,
avoiding star formation regions with bright supergiants and choosing
regions containing a great number of red giants  in order to diminish
the statistical error in the determination of the tip of red giant branch.
We used the fields S2 in NGC~891 (Fig.1), S1 in NGC~4144 (Fig.2), S1, S2
and S3 in NGC~4244 (Fig.3).

The luminosity function in the $I$ band of stars in each field has a jump
which corresponds to the beginning of the red giant branch (Fig.5).
Given the apparent  stellar magnitude of the tip of red giants branch,
one can measure the  distance to the galaxy by the TRGB
technique [41]. Simultaneously with the  determination of the distance the
average metallicity of red giants in the region of the galaxy being
investigated was measured.

It should be noted that due to the gradients of metallicity of red giants
along the radius of the galaxy, the local metallicity values that we
obtained cannot be regarded to be the metallicity of red giants of the
whole galaxy and use them for construction of global statistical
relationships of parameters of galaxies. This remark does not refer
to dwarf irregular galaxies where the gradient of metallicity along the
 radius of a galaxy is likely to be very insignificant. Using the red
giants of the ACS/WFC image beyond the border of the thick disk of
NGC~891 (Fig.1), we derived the value of $I_{TRGB}$ = 25.97, which
corresponds to ($m-M)= 29.96\pm0.08$, $D= 9.82\pm0.37$ Mpc) at [Fe/H] =
 $-$0.74.
The red giants of this galaxy have the highest metallicity of all three
galaxies. This is defined by the fact that the galaxy NGC~891 also has
the greatest mass preventing the outflow of metalls outside the border
of the galaxy. The measurements of the distance to NGC~891 were made
earlier on the basis of the luminosity function of planetary nebulae and
the method of the surface brightness fluctuations[42]. The mean
distance module obtained by these methods is $(m-M)=29.95\pm0.10$,
which corresponds to $D$ = 9.77 Mpc. The  discrepancy with our results
is very insignificant.

In the ACS/WFC field of the galaxy NGC~4144 we used red giants of thick disk
and halo outside the border of the main body of the galaxy. The tip of the
red giant branch is well noticeable at $I_{TRGB}=25.20$, which corresponds
to the distance modulus $(m-M)=29.30\pm0.10$ $(D=7.24\pm0.35$ Mpc)
at the mean metallicity of red giants [Fe/H] =$-$0.82. The distance to this
galaxy, $D=9.7$ Mpc was obtained earlier on the basis of photometry of the
brightest supergiants [43].
The difference between this result and our value can be explained by the
inclination of the galaxy, when only part of the galaxy is visible and
the calibration relations between the total luminosity of the galaxy and
the luminosity of the brightest stars may have a great error.

For NGC~4244 we obtained $I_{TRGB}=24.14$ and [Fe/H] = $-$1.62 for S1,
$I_{TRGB} = 24.20$ and [Fe/H] = $-$1.66 for S2 and $I_{TRGB}=24.15$
and [Fe/H] = $-$0.85 for S3. The average distance modulus
is ($m-M)=28.16\pm0.08$, which corresponds to ($D=4.33\pm0.16$ Mpc).
The discrepancies in the obtained metallicity values are due to
the fact that when calculating
TRGB and the metallicity of the red giants in S1 and S2 fields, we  used
stars at a maximum distance from the galactic plane, where a low--metallicity
population is observed and brigh AGB stars are absent. This improved
the accuracy of measuring the distance to the galaxy, and in the S3 field
stars were selected within the thick disk. The distance that we measured
is consistent, within the errors, with the distance $D=4.49$ Mpc obtained
by Karachentsev et al. [44] on the basis of photometry of the red giants of
the field S1.

\subsection{Distribution of stars over the body of galaxies}

The three galaxies being investigated are viewed nearby exactly edge-on,
and we could study of the number densities of different type stars
perpendicularly to the equatorial plane of the galaxy (Figs.6, 7, 8).
The principal difficulty we had  to overcome
when preparing the results was the scatter and discontinuity of the
fields of observations which we used to study the spatial distribution of
stars. When interpolating the results of the number density of stars of
different subsystems, we proceeded from the assumption that thick disks and
halos are of smooth, axially symmetric shape, without any tidal distortions.
It can be seen from Figs. 7 and 8 that the chaotic fluctuations of stellar
density near the equatorial plane of the galaxy disappear with increasing
distance from the plane of the galaxy, and the relationship ``number
density  - $Z$ coordinate'' assume an exponential form. This is
explained by the fact that at small distances from the equatorial plane
screening of stars by gaseous-dust matter of the galaxy decreases
essentially the luminosity of stars, and faint stars cease to be visible.
Besides, excess number density of stars is observed near the equatorial
plane.  Under these conditions the automated program FIND of DAOPHOT II
terminated recognition of closely spaced stars and does not include them
in the list of photometry.

The two causes mentioned lead to a fictitious decrease in the computed
star number density, which we observe in  Figs.7 and 8  at small distances
from the galactic plane. For the regions remote from galactic plane, the
exponential change of the stellar density, with no distorting factors being
present, points to a disk's character of stellar subsystems [45].

\subsubsection{NGC~891} As can be seen from Fig.1 the areas that we used
are located along the minor axis of the galaxy. The orientation of the image
of the ACS camera is not quite convenient for calculations of the star number
density along the $Z$ axis, but in return it allows the variation
of this density to be traced a great distance. The region of the image
near the galactic plane have considerable deficiency of red giants due
to the effects of screening the opposite parts of the disk and halo by the
body of the galaxy, therefore we observe a dip of number density of red
giants there. Similar but a smaller dip of density is observed for brighter
AGB stars as well. We excluded these regions absorption from the plot the
distribution of the number density of stars along the $Z$ axis to
visualize better the behavior of the number density of AGB and RGB stars
at the boundary of the thick disk and halo (Fig.6). The behavior of
the number density of stars of different types near the
galactic's plane will be  discussed in more detail below by the examples
of the galaxies NGC~4144 and NGC~4244 (Figs. 7 and 8).

The boundary of the thin disk in spiral galaxies is defined by the region
of spread of gaseous-dust clouds and young stars. In NGC~891 this boundary
is outside the interval in the $Z$ coordinate, which is presented on the
diagram in Fig.6. That is all AGB and RGB stars the distribution of
which is shown in Fig.6 belong to the thick disk and halo. AGB stars in the
disk of NGC~891 display a sharp drop in number along the $Z$ coordinate,
and outside the limits $Z$ = 6 kpc (Fig.6a) their number is small. The
number of RGB stars decrease essentially slower, and the point change of the
gradient of their member density, that is the boundary of the thick disks
becomes  noticeable (Fig.6b). Probably, the matter is that
massive galaxies viewed edge-on have a halo with a large stellar density
gradient of comparable with that of stellar density of the thick disk,
and the revealing of the bend point of the number density of stars in
such galaxies is an involved problem. The bend of the density gradient
in RGB stars is observed at $Z$ = 7.6 kpc (Fig.6b). By analogy with the
results for face-on galaxies [15] the bend point of the density of red
giants defines the boundary of the thick disk and the beginning of the more
extended halo. Having calculated the density gradient of RGB stars from
the inner part of the halo, we extrapolated the changes of stellar
density to zero values of the number density of stars. Thus, we computed
that the supposed size of the halo along the $Z$ axis is equal to 26.9 Kpc.
We chose the exponential character of the drop in the density of halo stars
by analogy with the thick disk. Fortunately, the field WFPC2 (S2) was found
exactly at the supposed boundary of the halo, which confirmed the
legitimacy of our extrapolation and permitted to refined the size
of the halo. Fig.6c shows the behavior of the number density of RGB
stars in the field S2. The stellar density diminishes along the $Z$
axis and falls to the background values at $Z$ = 23 kpc, which nearby
coincides with the result of our extrapolation as to the size of the halo.
 Thus, the measurements show that NGC~891 has a thick disk of 15 kpc
in thickness and a halo with a size of 46 kpc, along the $Z$ coordinate.
Proceeding from the assumption that the gradient of the decreasing of stellar
density along the major axis has the  same value (generally, it has smaller
value) as along the minor axis, it follows from thus assumption that the
halo is oblate at the poles of the galaxy. It is seen in Fig.1 that
assuming the found boundaries of the halo as a basis, we cannot inscribe
the round spherical halo into the dimensions of the galaxy, that is,
the halo of NGC~891 is an oblate ellipsoid in shape. Such
a shape is not unexpected since the oblateness of the halo at the poles is
likely to be also observed in the galaxy M~31 by the visible surface
distribution of bright red giants [46].

\subsubsection{NGC~4144} Fig.7 displays the plots of the
number density distribution of young stars, stars of intermediate
age (AGB) and old stars  (RGB). The  young stars in the ACS/WFC image are
distributed on the outer side of the thin disk  at a certain optical
depth, which is evidenced by the broading of branch of blue supergiants
(Fig.4). These stars has a maximum visual brightness and are slightly
affected by the brightness and inhomogeneity of the background near
the galaxy plane. This can be seen on the diagram of Fig.7, where the
the decrease of their number density in the distribution of stars
is not noticeable. Intermediate age stars (AGB) have a smaller gradient
of number density, as compared to young stars have, and extend to larger
distances from the galactic plane than young stars. The most numerous
stars --- red giants have the smallest gradient of the number density.
At distances $Z$ = 2.4  kpc from the galactic plane, NGC~4144 shows
a change of the number density gradient of RGB stars as in the galaxy
NGC~891. We assume this point to be the boundary of the thick disk.
At larger distance from the galactic plane, in the halo region,
extrapolation of the exponential decreasing of the number density of stars
makes it possible to compute the boundary of the halo ($Z$ -- 5.4 kpc).
The same as for the galaxy NGC~891, the halo of NGC~4144 has an oblate
shape at the poles of the galaxy.

\subsubsection{NGC~4244} Although the galaxy NGC~4244 resembles NGC~4144
both in morphology and stellar composition, there are some differences
of its parameters of outer stellar structures. Fig.8 shows plots of
the number density distribution of stars of different age. The same as in NGC~4144,
young stars of NGC~4244 concentrate in the most narrow equatorial region.
Intermediate age stars have a wider distribution, while old stars have
a minimum gradient of the number density and occupy the  most extended
region. The difference between the galaxies is that the plot (Fig.8a) shows
no bend point of the number density of red giants in NGC~4244, which would
correspond to the boundary between the thick disk and the halo. To clear up
the point whether  a halo exist or not in this galaxy we have studied the
distribution of red giants in the fields S1 and S2 of the WFPC2 (Fig.3).
The field S1 does not  fall outside the thick disk of the NGC~4244 and no
changes in the density gradient are noticeable in the number density
distribution of red giants (Fig.8). However, in the field S2 the distribution
of RGB stars shows a sharp change of the number density of stars along the
$Z$ axes. To the $Z$ = 2.7 kpc the number density gradient corresponds to
that of the field ACS/WFC, but at greater distances from the galactic plane
the number density gradient undergoes an clean-cut change, that is, the
boundary of the thick disk is observed (Fig.8b). Transition from the thick
disk to the halo occurs at small values of stellar number density. The
boundary of the halo ($Z$ = 8 kpc) can be estimated approximately because of
lack of data, but the existence of a halo is of  no doubt. Thus, the galaxy
NGC~4244 has a thick disk 5.4 kpc in thickness and a halo with low surface
density, having a thickness of 16 kpc.

\section{Observational model of stellar structure of a spiral galaxy}

On the basis of earlier studies of the stellar population of three spiral
galaxies, we present a model of stellar structure of a spiral galaxy [15].
We considered in this model the variation of the number density of
different-type stars along the radius of the galaxy from bulge to halo. It
was shown that the boundaries of the thin disk are well defined by the region
of distribution of young stars. The boundary of the thick disk and halo
are defined by the bend point of the number density gradient of old
stars -- red giants. However, due to the lack of data, we failed to present
the distribution of different type stars perpendicular to the galactic
plane. But it is exactly such distribution give an conception of the three
-- dimensional shape of galaxies. In the present paper we took as a basis
this, the already mentioned model, but added to it our new results on the
investigation of the distribution of stars in the galaxies along the $Z$
axis (Fig.9).

Qualitatively, the distribution of stars both along the radius of the galaxy
and along the $Z$ axis are alike. Young stars occupy the region of the
thin disk, intermediate age stars form a structure of a larger size, while
old stars form the thick disk and halo. The results of stellar structure
of three massive irregular galaxies, IC~10, M~82, NGC~3077, where we have
found the existence of thick disks and halos fall within this model.
A similar at the first sight, model of  composition of spiral galaxies,
one can find in any book. The differences are in that the models of stellar
structure of spiral galaxies presented earlier are constructed to a
considerable degree on the  basis of the  spatial structure of the
only Galaxy or on speculative assumption without references to particular
investigations. Presenting results of investigation of the stellar
structure of a great number of spiral galaxies, which are consistent
with one another, we hope the observational data that we have obtained
will facilitate creation of theoretical models of origin and evolution
of spiral and irregular galaxies.

\section{Discussion}
Resolution of galaxies into stars will make it possible to investigate
the spatial distribution of different types stars, both young and
intermediate age  star or old stars, in galaxies. The possibility of
selection of stars by their types makes the method of counting stars
basically different from the method of surface photometry since it
provides a  possibility of dividing stars of different luminosity and age
but of similar colour (for instance, AGB and RGB stars). A second
distinction of the method of counting stars is that it makes possible
to reveale spatial stellar structures of very low surface brightness,
which cannot be done by the method of surface photometry. We have used
the indicated advantages to clear up the composition and extent of stellar
structures of three edge-on spiral galaxies. Under investigations of these
galaxies, NGC~891, NGC~4144 and NGC~4244, it was ascertained that although
morphologically all of them have the same structure: a thin and thick disks
and halo, but the relative dimensions of the thick disk and halo change
within a wide range. Nor does the surface brightness of the halo remain
constant at border with the thick disk. Whereas in NGC~891 the halo extends
to the size $Z$ = 23 kpc and has hight surface brightness, in NGC~4244
the halo is three times as small and has a low surface brightness which
turned out to be barrier to its detection by the method of surface
photometry [36]. Taking into consideration the accidental choice of
the galaxies and also the fact we already revealed earlier thick disks
and halos in three spiral galaxies one may consider with high probability
that the thick disks and halos composed from old stars are necessary
components of spiral galaxies. Apart from the galaxies discussed in
the present paper, the edge-on galaxies IC~2233, IC~5052 and NGC~4631
have a quits similar structure, which chances statistical significance
of the results that we have obtained. It is natural that the process
of interaction of galaxies can distort the stellar structure, and
our conclusions are inapplicable to such galaxies.

\section{Conclusions}
Based on the method of number density calculation of stars, an investigation of
stellar structures of three spiral galaxies viewed edge-on has been carried
out. The study of the stellar population of different age has made it
possible to obtain the following results:

a) For the first time, thick disks and halos constituted mainly from red
giants have been revealed in the galaxies NGC~891, NGC~4144 and NGC~4244.

b) The difference between the stellar density gradients of the thick disk
and halo, which we found earlier, has been confirmed, which allows their
spatial sizes to be objectively determined.

c) It has been established that halos of spiral galaxies have a shape
 oblate at the galaxy poles, which is likely to point to its rotation

d) Based on the photometry of red giants the distances of the galaxies
NGC~891 and NGC~4144 have been measured for the first time.

e)  On the basis of the results obtained a model of stellar structure of
spiral galaxies is presented. The use of the method of counting of stars
has shown that measurements of stellar number density can be successfully
employed for revealing of stellar subsystems and establishment of
boundaries betwen thick disks and halos.

This work was done under the support of RFBR grant 03--02--16344.

\bigskip
{\large \bf References}
\bigskip

1. Vallenari A., Bertelli G., Schmidtobreick L., Astron. Astropys., 361,
73, 2000.

2. Prochaska J.X., Naumov S.O., Carney B. W., McWilliam A., Wolfe A. M.,
Astron. J., 120, 2513, 2000.

3. Chiba M., Beers T.C., Astron. J, 119, 2843, 2000.

4. Zoccali M., Renzini A., Ortolani S., Greggio L., Saviane I., Cassisi S.,
   Rejkuba M., Barbuy B., Rich R.M., Bica E., astro-ph/0210660, 2002.

5. Williams B.F., Mon. Notic. Roy. Astron. Soc., 331, 293, 2002.

6. Rowe J.F., Richer H.B., Brewer J.P., Grabtree D.R., astro-ph/0411095,
   2004.

7. Schombert J.M., Bothum G.D., Astron. J., 93, 60, 1987.

8. Dalcanton J.J., Bernstein R.A., Astron. J., 124, 1328, 2002.

9. Minniti D., Zijlstra A. A., Astrophys. J, 467, 13, 1996.

10. Minniti D., Zijlstra A. A., Alonso M. V., AJ, 117, 881, 1999.

11. Tikhonov N.A., Dissertation, St.-Peterburg, Russia, 2002.

12. Tikhonov N.A., accepted to Astronomy Reports,  2005a.

13. Tikhonov N.A., submitted to  Astronomy Reports, 2005b.

14. Drozdovsky I., Tikhonov N., Schulte-Ladbeck R., "The outer stellar
    edges of irregular galaxies: IC10 and LeoA", STScI May, 2003.

15. Tikhonov N.A., Galazutdinova O.A., Drozdovsky I., Astron. Astropys.,
    431, 127, 2005.

16. Gonzalez J.J., PhD thesis Univ. California, Santa Cruz, 1993.

17. Cantiello M., Raimondo G., Brocato E., Capaccioli M., Astron. J., 125,
    2783, 2003.

18. Pasquini L., Bonifacio P., Randich S., Galli D., Gratton R.G., astro-
    ph/0407524, 2004.

19. Howk J.C., Savage B.D., Astron. J., 114, 2463, 1997.

20. van der Kruit  P., Searle L., Astron. Astropys., 95, 116, 1981.

21. Garcia-Burillo S., Guelein M., Cernicharo J.J., Dahlem M., Astron.
    Astropys., 266, 21, 1992.

22. Swaters R.A., Sancisi R. and van der Hulst J.M., Astrophys. J, 491, 140,
    1997.

23. Flaternali F., Oosterloo T., Recycling intergalactic and interstellar
    matter IAU Symposium Series, v. 217, astro-ph/0310799, 2004.

24. Flaternali F., Oosterloo T., Sancisi R., Swaters R., astro-ph/0410375,
    2004.

25. Martin M.C., Astron. Astrophys. Suppl. Ser., 131, 77, 1998.

26. Xilouris E. M., Alton P. B., Davies J. I., Kylafis N. D., Papamastorakis
    J., Trewhella M., Astron. Astrophys., 331, 894, 1998.

27. Spitzer L., Physical Processes in the Interstella Medium, New York,
    Wiley-Interscience, (p.162), 1978.

28. Hunter D.A., Gallagher J.S. III, Astron. J., 90, 1789, 1985.

29. Alonso-Herrro A., Knapen J.H., Astron. J., 122, 1350, 2001.

30. Garcia-Ruiz I., Sancisi R., Kuijken K., Astron. Astrophys., 394, 769,
    2002.

31. Swaters R. A., van Albada T. S., van der Hulst J. M., Sancisi, R.,
    Astron. Astropys., 390, 829, 2002a.

32. Swaters R. A., Balcells M., Astron. Astrophys., 390, 863, 2002b.

33. Hummel E., Sancisi R., Ekers R. D., Astron. Astrophys., 133, 1, 1984.

34. Olling R.P., Astron. J, 112, 457, 1996.

35. Kodaira K., Yamashita T., Publ. Astr. Soc. Jap., 48, 581, 1996.

36. Fry A.M., Morrison H.L., Harding P., Boroson T. A., Astron. J., 118,
    1209, 1999.

37. Stetson P.B., Users Manual for DAOPHOT II, 1994.
38. Dolphin A.E., Publ. Astr. Soc. Pacif, 112, 1383, 2000.

39. Holtzmann J.A., Hester J.J., Casertano S., Publ. Astr. Soc. Pacif.,
    107, 156, 1995a.

40. Holtzmann J.A., Burrows C.J., Casertano S., Hester J.J., Trauger J.T.,
    Watson A.M., Worthey G., Publ. Astr. Soc. Pacif., 107, 1065, 1995b.

41. Lee M.G., Freedman W.L., Madore B.F., Astrophys. J., 417, 553, 1993.

42. Ferrarese L., Mould J.R., Kennicutt R.C et al., Astrophys. J, 529, 745,
    2000.

43. Karachentsev I.D., Drozdovsky I.O., Astron. Astrophys. Suppl. Ser., 131,
    1, 1998.

44. Karachentsev I.D., Sharina M.E., Dolphin A.E., Grebel E.K., Geisler D.,
    Guhathakurta P., Hodge P.W., Karachentseva V.E., Sarajedini A., Seitzer
    P., Astron. Astrophys., 398, 467, 2003.

45. de Grijs R., van der Kruit P.C., Astron. Astrophys. Suppl. Ser., 117,
    19, 1996.

46. Zucker D., Kniazev A., Bell E., Martinez-Delgado D., Grebel E., Rix H,W.,
    Rockosi C., Holzman J., Walterbos R. et al., Astrophys. J., 612, L117,
    2004.

47. Schlegel D. J., Finkbeiner D.P., Davis M., Astrophys. J., 500, 525, 1998.

\newpage
\renewcommand{\tabcolsep}{2pt}
\begin{table}
\footnotesize
\caption{Galaxies Data}
\begin{tabular}{lccccccccccc}\\ \hline
\multicolumn{1}{c}{Name}&
\multicolumn{1}{c}{$V_r$}&
\multicolumn{1}{c}{$a^{\prime}\times b^{\prime}$}&
\multicolumn{1}{c}{$B_t^0$}&
\multicolumn{1}{c}{$Classification$}&
\multicolumn{1}{c}{$A_b$}&
\multicolumn{1}{c}{$A_v$}&
\multicolumn{1}{c}{$A_i$}&
\multicolumn{1}{c}{$i$}&
\multicolumn{1}{c}{$M-m$}&
\multicolumn{1}{c}{$M_{abs}$} &
\multicolumn{1}{c}{} \\ \hline \\
NGC 891 & 528  & 13.5$\times$2.5& 9.37&SA(s)b sp HII  & 0.280& 0.215&
0.126& 90& 29.96& $-$20.59&\\
NGC4144 & 265  &  6.0$\times$1.3& 11.10&SAB(s)cd sp HII& 0.065& 0.050&
0.290& 84& 29.30& $-$18.20&\\
NGC4244 &244   & 19.4$\times$2.1& 9.28&SA(s)cd:sp HII  & 0.090& 0.069&
0.040& 90  &  28.16& $-$18.88&\\
\hline
& & & & & & & & & & & \\
\multicolumn{12}{l}{The Galactic extinction correction is by Schlegel et
al.(1998).}\\

\multicolumn{12}{l}{The inclination is taken from LEDA.}\\

\multicolumn{12}{l}{Values of the ($M-m$) and $M_{abs}$ have obtained in this
paper.}\\
\end{tabular}
\end{table}

\begin{table}
\footnotesize
\caption{Observational log of HST.}
\renewcommand{\tabcolsep}{4pt}
\begin{tabular}{lccccccr}\\ \hline \hline
\multicolumn{1}{c}{Galaxy}&
\multicolumn{1}{c}{Region}&
\multicolumn{1}{c}{Date}&
\multicolumn{1}{c}{Band}&
\multicolumn{1}{c}{R}&
\multicolumn{1}{c}{Exposure}&
\multicolumn{1}{c}{ID}&
\multicolumn{1}{r}{N$_{stars}$ }\\
                            \hline\\
NGC891 & S1   & 2003-02-19 & F814w&  2.12  &  2$\times$2620+2472 & 9414&
108970\\
       &      & 2003-02-19 & F606w&  2.12  &  2$\times$2620+2472 & 9414&\\

       & S2   & 2003-02-20 & F814w&  7.94  &  2$\times$1000  & 9676& 900\\

       &      & 2003-02-20 & F606w&  7.94  &  2$\times$400+4$\times$500 &
 9676&\\
NGC4144& S1   & 2003-12-08 & F814w&  0.98  &  350  & 9765& 63605\\
       &      & 2003-12-08 & F606w&  0.98  &  338  & 9765&\\
NGC4244& S1   & 2001-06-11 & F814w&  3.77  &  600  & 8601& 9680\\
       &      & 2001-06-11 & F606w&  3.77  &  600  & 8601&\\
       & S2   & 2001-06-30 & F814w&  2.09  &  6$\times$500  &9086& 3945\\
       &      & 2001-06-30 & F606w&  2.09  &  6$\times$500  &9086&\\
       & S3   & 2003-11-12 & F814w&  0.33  &  350  & 9765&115990\\
       &      & 2003-11-12 & F606w&  0.33  &  338  & 9765&\\

\hline
\end{tabular}
\end{table}

\begin{figure}
\centerline{\psfig{figure=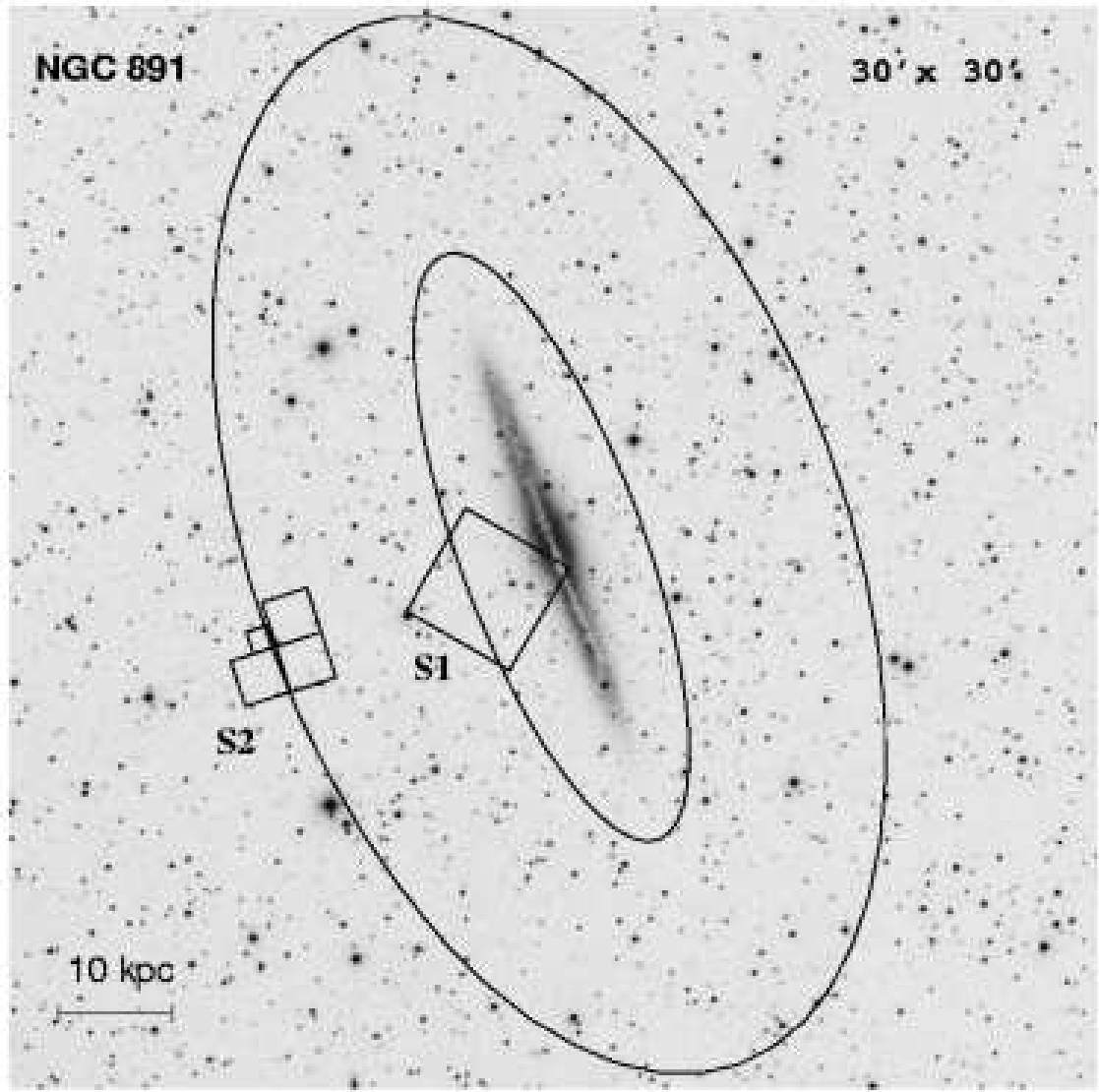,width=15cm,angle=0}}
\caption{DSS-2 image of the galaxy NGC~891 with the S1 (ACS/WFC)
and S2 (WFPC2) footprint overlaid.
The inner ellipse shows the boundary of the thick disk, the outer one
shows that of the halo.}
\end{figure}
\begin{figure}
\centerline{\psfig{figure=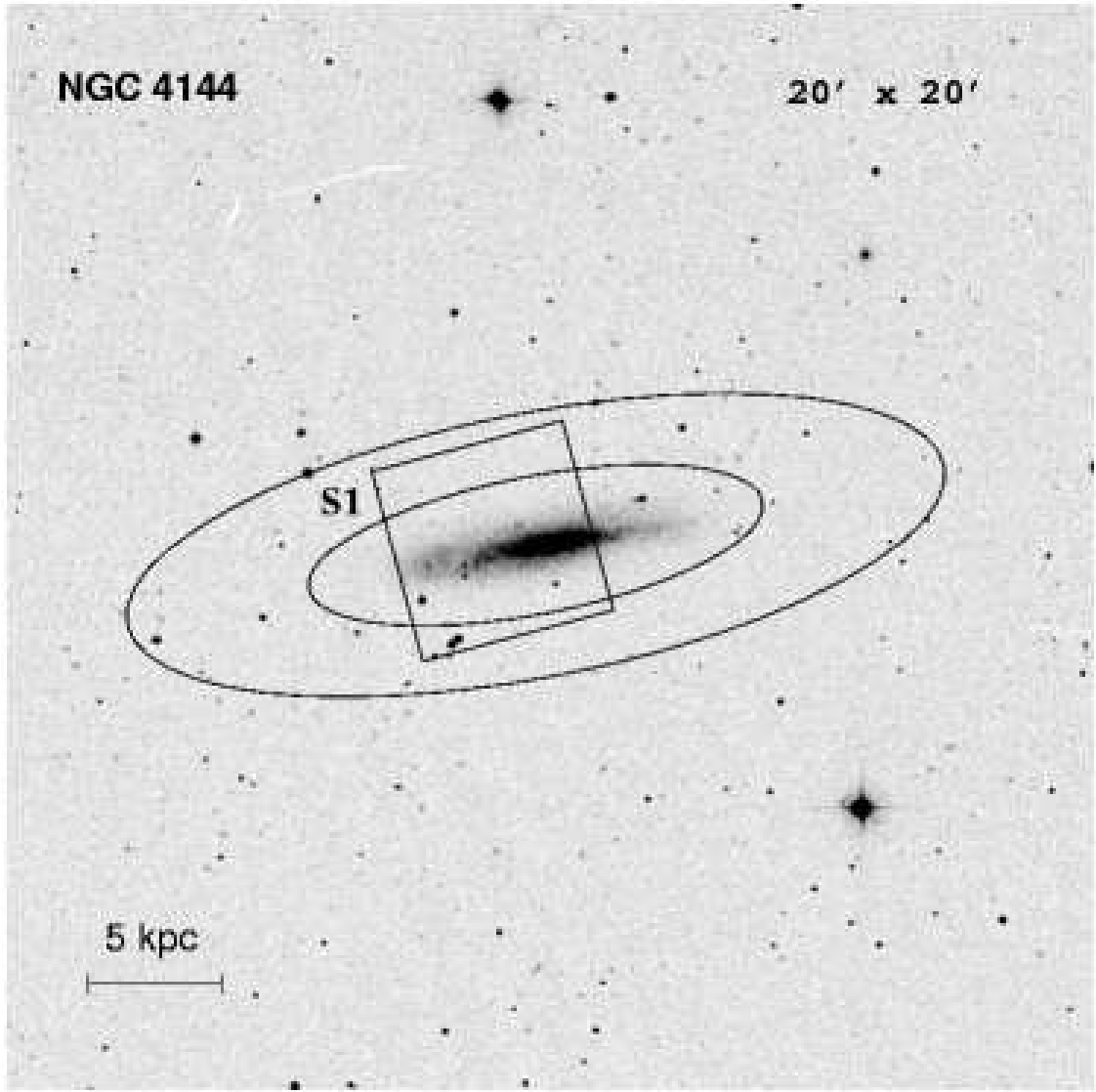,width=15cm,angle=0}}
\caption{Same as in Fig.1 for the galaxy NGC~4144.           }
\end{figure}
\begin{figure}
\centerline{\psfig{figure=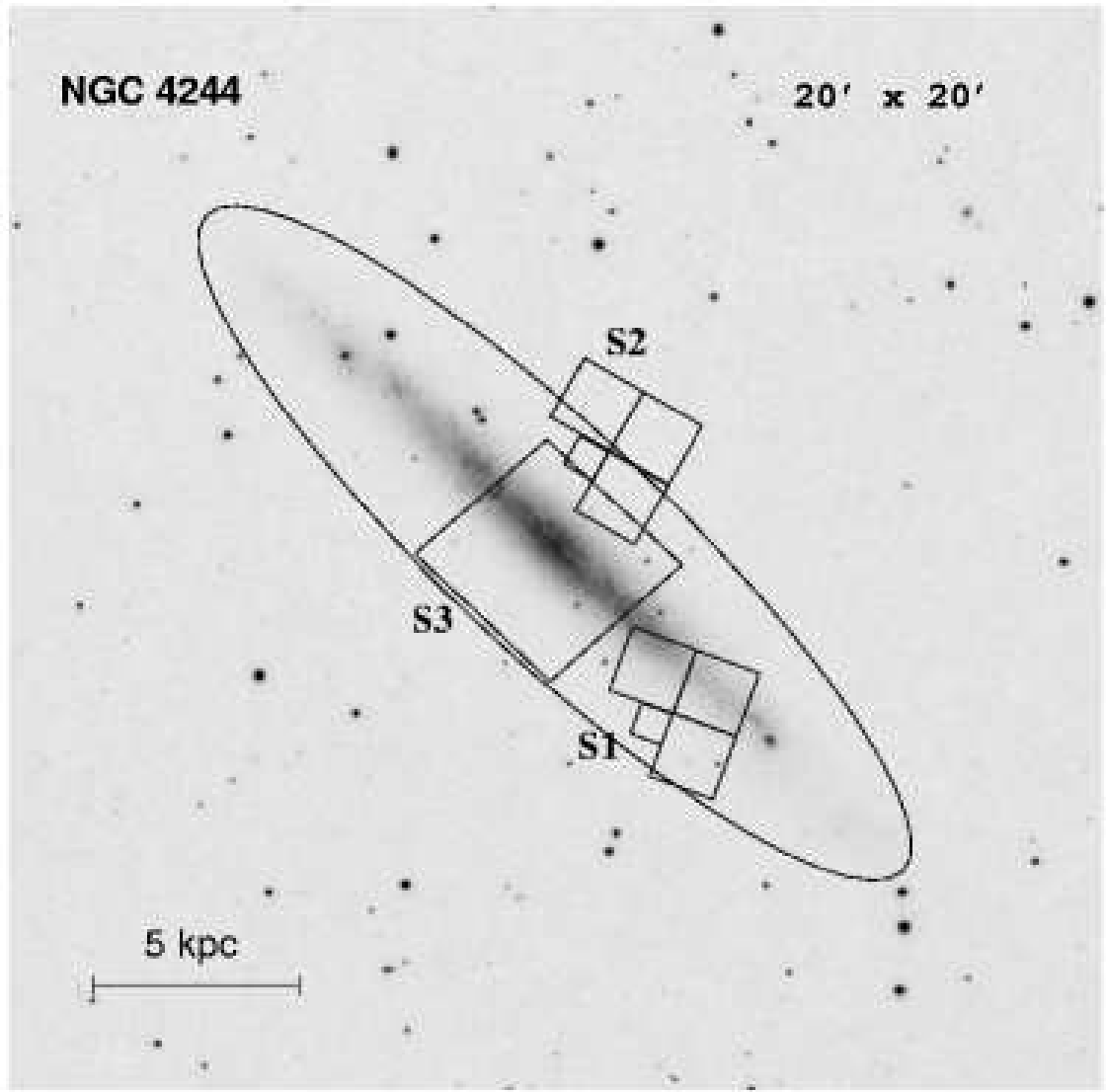,width=15cm,angle=0}}
\caption{Same as in Fig1 for the galaxy NGC~4244. Only the boundary of the
thick disk is marked.}
\end{figure}
\begin{figure}
\centerline{\psfig{figure=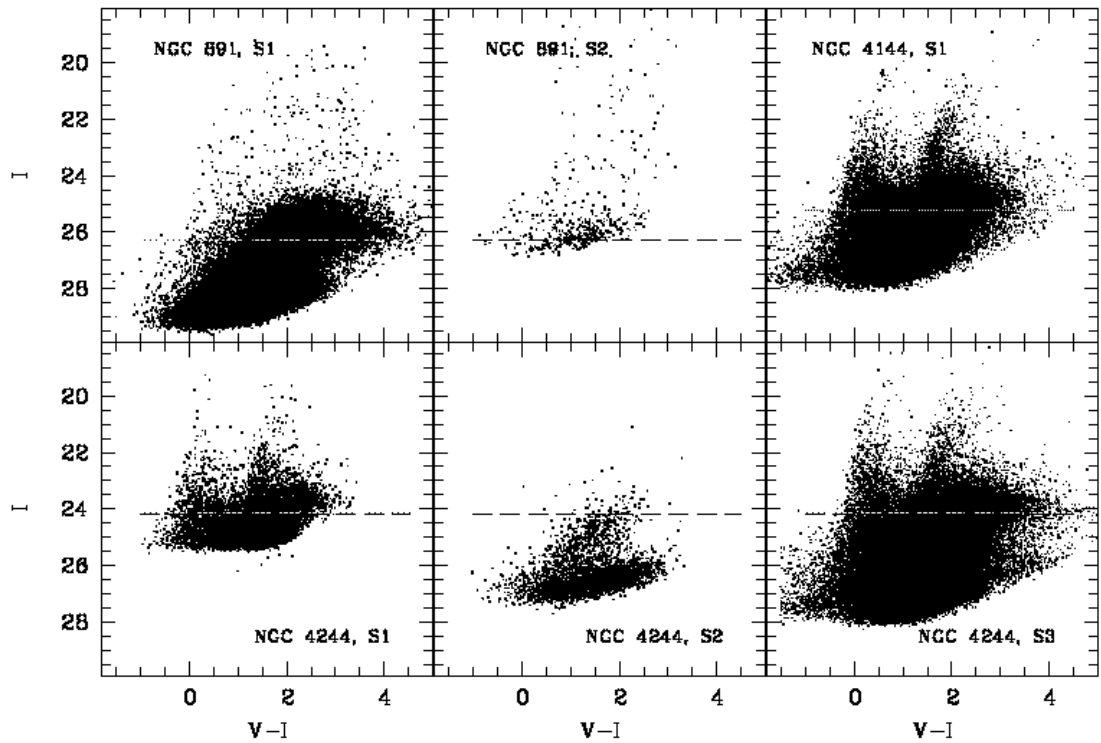,width=15cm,angle=-90}}
\caption{Color--Magnitude diagram for all fields of the studied galaxies.
The dotted line marks the position of the tip of the red giant branch
(TRGB).}
\end{figure}
\begin{figure}
\centerline{\psfig{figure=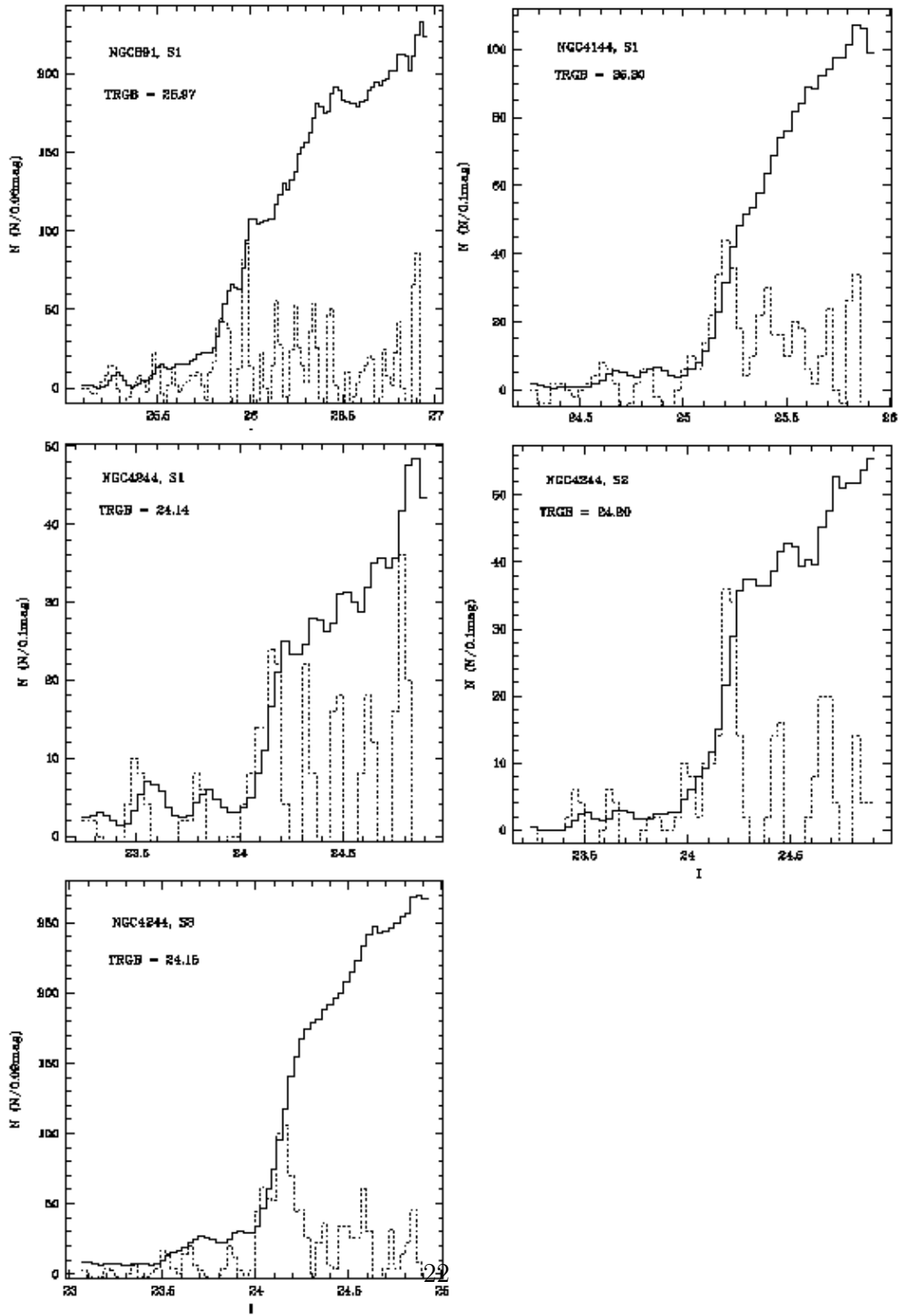,width=15cm,angle=0}}
\caption{I band luminosity function of the stars in fields being
investigated in. The sharp change in the number of stars corresponds to the
beginning of the red giant branch, which is used to measure the distances
(TRGB method).}
\end{figure}
\begin{figure}
\centerline{\psfig{figure=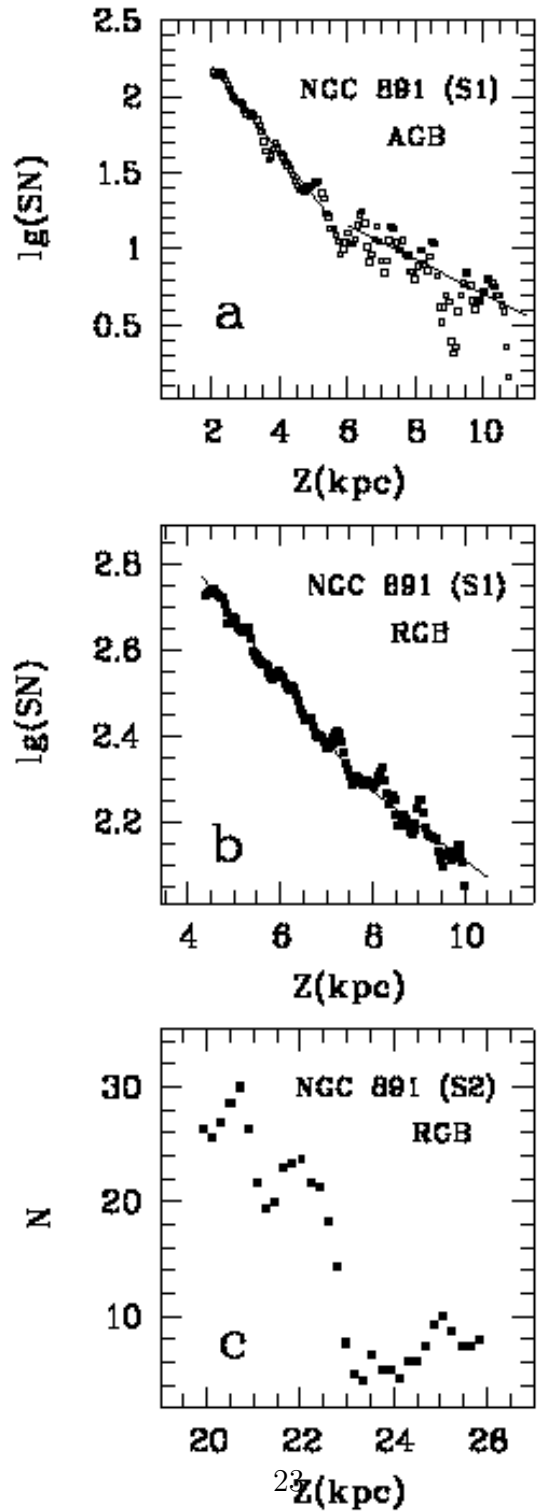,width=9cm,angle=-90}}
\caption{ Number density distribution of AGB (a) and RGB (b) stars
of NGC~891 at the border of the thick disk (field S1) and RGB stars
(c) at the border of the halo (field S2). At $Z$ = 7.6 kpc a bend of
the number density of RGB stars corresponds to the transition from the
thick disk to the halo. In the field S2 (c) after $Z$ = 23 kpc number
density distribution of stars is of flat character and corresponds
to the level of background stars.}
\end{figure}
\begin{figure}
\centerline{\psfig{figure=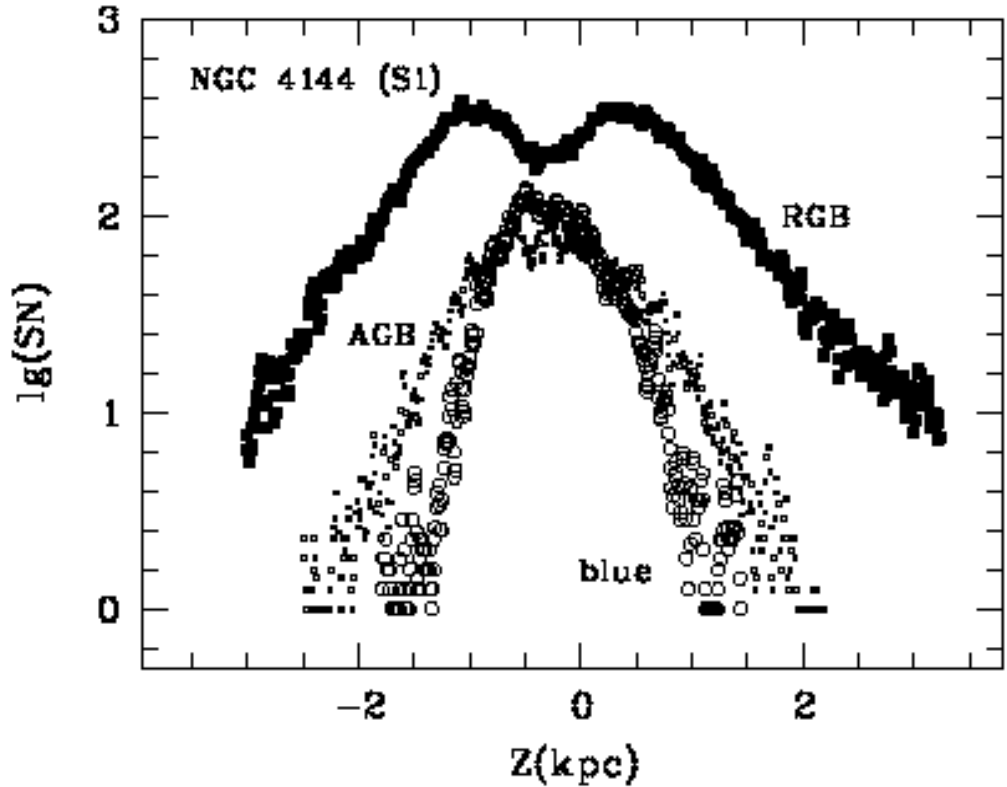,width=15cm,angle=-90}}
\caption{Distribution of the number of stars of different age perpendicular
to the plane of galaxy NGC~4144. Open circles are young stars. Dots are the
stars of intermediate age (AGB). Filled circles mark old stars (RGB). At
$Z$ = 2.4 kpc a change in the gradient of the number density of RGB stars
corresponds to the transition from the thick disk to the halo.}
\end{figure}
\begin{figure}
\centerline{\psfig{figure=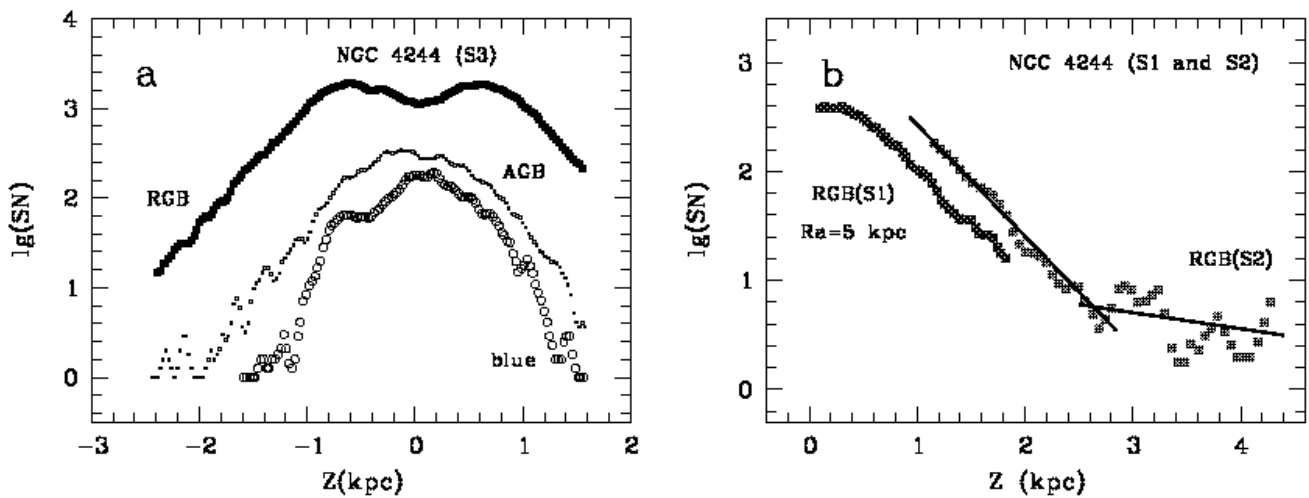,width=18cm,angle=-90}}
\caption{Same as in Fig.7 for the galaxy NGC~4244. It can be seen that
the boundary of the thick disk is outside the field S3 (b). The
distribution of old (RGB) stars in the fields S1 and S2 is perpendicular
to the galaxy plane. At $Z$ = 2.7 kpc transition from the thick disk to
the halo is seen.}
\end{figure}
\begin{figure}
\centerline{\psfig{figure=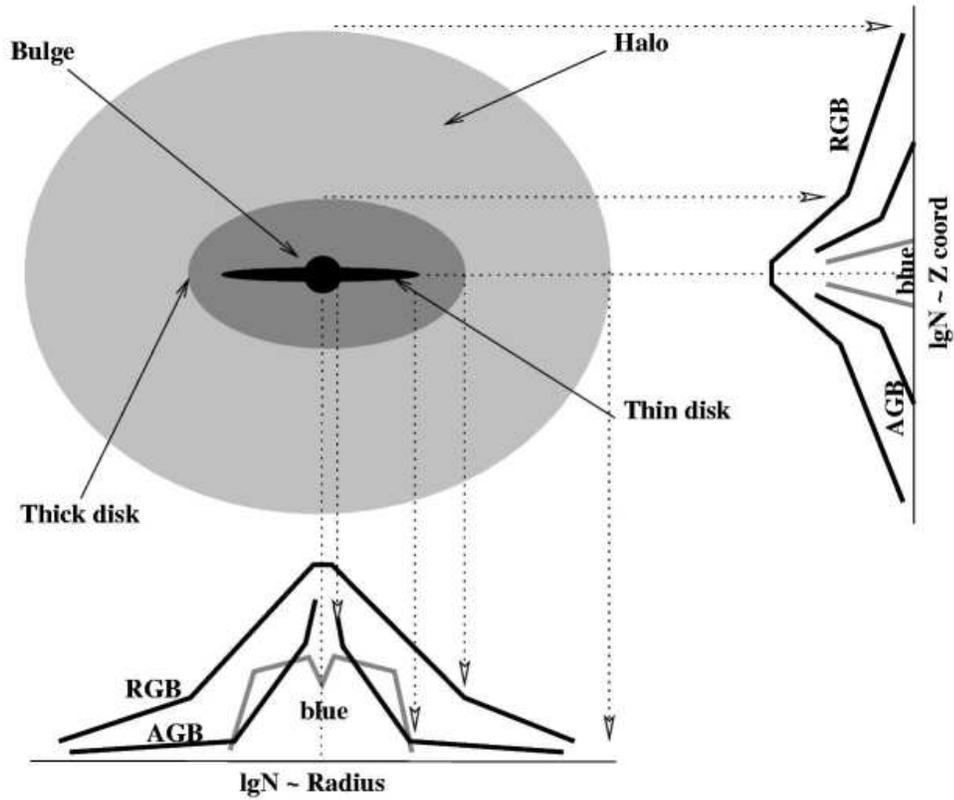,width=15cm,angle=0}}
\caption{Three--dimensional model of the number density distribution
of the stellar population along the radius and $Z$ axis in spiral galaxies.
Number densities of different type stars are given in relative units. The
absolute sizes of the halo along the $Z$ coordinate change from 8
to 25 kpc in each individual galaxy.}
\end{figure}

\end{document}